Research Article

# Quantifying and Predicting Pitch Sequence Effectiveness Using Visual Deception Metrics

**Joe DiGiovanni, Justin Kurland**

## Abstract

Pitcher performance is influenced not only by individual pitch characteristics, but also by the relationship and sequencing between consecutive pitches. In this study, we construct a sequence-level framework using MLB Statcast data from 2021–2025 to quantify the value of pitch deception with two geometric features: Decision Angular Separation (DAS) and Average Angular Separation (AAS). These metrics measure visual similarity of back-to-back pitches from the batter's perspective at a hitter's swing "decision point" during the flight of the ball from the pitcher's hand to the plate. We introduce Swinging Strike Lift (SwStr Lift), a residualized outcome isolating sequencing value beyond player identity, pitch type, location, and count.

Across more than 740, 936 two-pitch sequences, we find that tighter tunnels (low DAS/AAS) produce significant behavioral changes, for well-tunneled FB→BR (fastball→breaking-ball) sequences in the Shadow/Chase regions. This elevated swing propensity is consistent with swing induction as a primary behavioral pathway underlying the observed gains in swinging strikes. Our predictive modeling confirms the importance of tunneling, as adding DAS/AAS to gradient-boosted models yields substantial relative improvements in out-of-sample $R^2$ (e.g., +38% for FB→BR and +163% for FB→OS [fastball→off-speed]). These results demonstrate that tunneling generates a measurable and behaviorally coherent sequencing advantage, offering a tractable, perceptually grounded tool for evaluating pitcher deception which informs pitch design and game strategy.

Quintile analyses confirm the effect is not confined to extreme tunnels; ICE plots show model-predicted lift declining sharply as DAS increases.

## 1 Introduction

### 1.1 Motivation and Background

Baseball is a game of sequential strategy. Individual pitch outcomes do not exist in isolation; a batter's approach is shaped by the pitch they saw before, setting the stage for the next pitch. Advances in baseball analytics have led to sophisticated measures of individual pitch quality such as velocity, spin rate, and horizontal/vertical movement, but there remains a critical gap: the marginal effect of pitch sequencing. In other words, how does the relationship between consecutive pitches alter the outcome in ways not captured by pitch-level metrics alone?

The importance of sequencing has long been acknowledged in coaching and scouting circles, often described informally as the art of “keeping hitters off balance.” Yet quantitative evidence isolating sequencing effects is relatively sparse compared to other dimensions of sabermetrics. Our work addresses this gap by proposing and empirically testing metrics of visual deception between consecutive pitches, termed Decision Angular Separation (DAS) and Average Angular Separation (AAS), and by introducing a novel outcome variable, Swinging Strike Lift (SwStr Lift), designed to measure the added value of sequencing above pitch-level expectations.

Foundational sabermetric texts such as Albert and Bennett (2003) emphasize that subtle, context-driven factors can substantially alter performance outcomes, which is the reason this study focuses on sequencing.

### 1.2 Literature Review

A substantial body of work evaluates contributions that a season line hides. Baumer (2009) used a Monte Carlo simulation of real players and found that baserunning typically changes a team’s run scoring by less than ±25 runs per season. A lineup of the best baserunners could gain about 70 runs, and a lineup of the worst could lose about 55. Sequencing is the same kind of effect: real, modest in a typical sample, and easy to miss in a raw total.

Kaplan (2006) decomposed offensive performance into a between-team component and a within-team individual component. Most of the variance sat with the hitter. Pitching needs the same split. A whiff can belong to the pitch itself or to the pitch that came before it.

Game theory has been used on the same problem. Weinstein-Gould (2009) tested pitch mixing on the first pitch of a plate appearance. That mix was consistent with an optimal mixed strategy for the result of the first pitch, and it was not consistent with optimal play for the outcome of the plate appearance. Unpredictability is a real target, and it is not automatic. Other work estimates effects that standard totals mix together. Acharya et al. (2008) improved park-factor estimates so that a run-scoring environment is not mistaken for offensive skill. Brill, Deshpande, and Wyner (2023) use a Bayesian hierarchical model to separate a time-through-the-order penalty from the way a pitcher changes across a game.

Albert (2016) applies Bayesian random-effects models to component rates such as strikeouts and home runs, then combines the components. Those predictions of later batting and pitching performance beat shrinking the aggregate statistic alone. Piette and Jensen (2012) estimate fielding ability across seasons, another skill that a season line does not show by itself.

The same standard shows up outside baseball. Spilker and Ötting (2024) found that playing professional darts without a crowd during COVID-19 changed performance, and that the change was not the same for every player. Streak studies are a warning about reading a sequence off a raw rate. Steeger, Dulin, and Gonzalez (2021) found that NHL winning and losing streaks were consistent with a run of independent games. Ding, Cribben, Ingolfsson, and Tran (2025) found no hot hand for playoff goalies, and some evidence that a recent string of saves made the next save less likely. A recent success is not, by itself, a sequencing effect. SwStr Lift is a residual for that reason.

Pitch framing is a closer baseball example. Deshpande and Wyner (2017) use a hierarchical Bayesian model to estimate how much a catcher changes the chance that a taken pitch is called a strike, after the pitch, the pitcher, the batter, and the umpire are accounted for.

### Pitch Tunneling and Arsenal-Level Models

Pitch tunneling is the idea that different pitch types can look alike for part of their flight. Sayre, Pavlidis, Judge, and Long (2017) put numbers on it. A league-average fastball reaches the plate in about 400 milliseconds, and they place the batter's point of no return 175 milliseconds before contact, 23.8 feet from home plate. They then measure how far apart consecutive pitches are at release, at that tunnel point, and at the plate.

Roegele (2014) had already tested the performance claim, at a decision point farther from the plate than 23.8 feet. Consecutive pitches of different types that nearly overlap when the swing must start, and then separate before they reach the plate, produced higher swinging-strike rates in both seasons he studied. Sayre et al. (2017) treat that result as the reason to publish the distances. Gray (2017) argues that one distance at one depth is not enough. Batters pick up optical information continuously, including the rate at which the ball's image expands, so two pitches that meet at the tunnel point can still be told apart earlier in the flight.

Later public work grades a pitch inside the rest of the repertoire. Lambert (2024) builds what he calls an arsenal effect into Driveline's Stuff+ model: a breaking ball or off-speed pitch is compared with the velocity and movement of that pitcher's primary fastball, split by batter handedness. Sutton-Brown (2025) does the comparison from the batter's view for Baseball Prospectus. Pitch Type Probability is the chance the hitter could name the pitch from its path up to the decision point and the count. Movement Spread and Velocity Spread describe how wide the remaining movement and velocity possibilities are. Surprise Factor is how unusual the actual movement is inside that mixture.

These arsenal grades describe a repertoire. They are not a forecast for the next pitch in a specific sequence, and they do not separate that sequence from the quality of the pitch itself. A high whiff rate on a tunneled slider can still be the slider's velocity and break.

## 1.3 Our Contribution and Hypotheses

In this study, we contribute to the sabermetric literature by isolating and predicting the sequencing advantage created by consecutive pitch sequences. A common misconception is that tunneling is the act of two different pitches appearing identical for as long as possible during flight before maximizing the Euclidean distance between them once they reach the plate. This paper challenges that assumption. To do so, we develop two new geometric features that capture the *visual separation* of consecutive pitches from the batter's perspective:

- Decision Angular Separation (DAS) measures the angular difference between the trajectories of two consecutive pitches at the decision point. We place that point at 24 feet from home plate, rounding the 23.8-foot tunnel point in Sayre et al. (2017). This is the depth at which a batter must begin the swing, so it is a natural point of perceptual evaluation (Figure 1). A smaller DAS means the two pitches looked more alike there, which raises the chance of a swing and miss.
- **Average Angular Separation (AAS):** extends this idea by calculating the *average angular difference* between pitch trajectories from the pitch release point to the decision point. While DAS reflects deception at the instant of commitment, AAS reflects the sustained visual relationship between pitches up to such commitment.

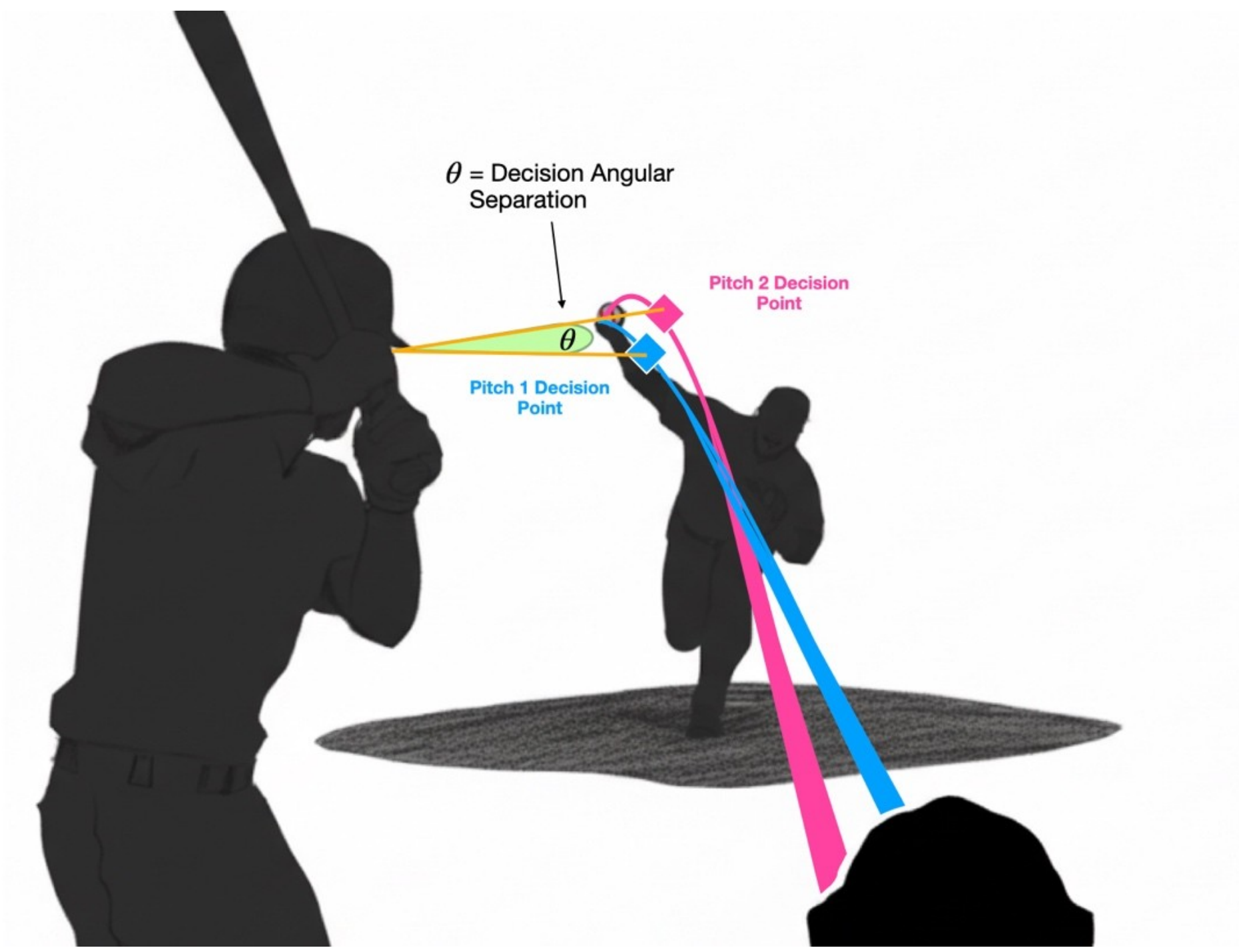

**Figure 1.** A visual depiction of tunneling with DAS in a two-pitch sequence.

Both metrics are derived using a simple geometric framework: treating the batter's eyes as the origin, each pitch trajectory is represented as a vector, and the angular separation θ between two pitches is given by the arccosine of the normalized dot product of their directional vectors. This aligns with the batter's perceptual challenge of distinguishing two moving objects on converging or diverging paths.

With these tools, we can frame and test two central hypotheses:

- **H1:** Sequences with lower DAS and AAS values (i.e., tighter tunnels) produce significantly higher SwStr Lift, controlling for pitch type, location, and count.

- **H2:** The primary mechanism for this effect is swing induction, as deceptive sequences increase the probability of swings at pitches in the Shadow and Chase zones, where batters are selective.

By introducing these geometric deception metrics and a sequence-specific predictive framework, this work seeks to offer actionable insights for baseball strategy, coaching, and player evaluation.

### 1.4 The Geometry of Pitch Deception's Relation to the Platoon Advantage

Before formalizing our predictive framework, it is necessary to establish the geometric basis of the "tunneling" hypothesis. Traditional analysis often assumes that two pitches with similar release points but largely different plate locations will look alike to the batter and induce a swing. However, this ignores the critical variable of the observer's perspective.

To illustrate this, we examine a particular two-pitch sequence (Fastball followed by Slider) thrown by Pirates RHP Paul Skenes, modeled in 3D space from two distinct vantage points (Figure 2). Although the same sequence produces only small quantitative differences in DAS across RHH/LHH viewpoints, the figure illustrates why evaluating tunneling from the batter's optical origin is required conceptually: perception depends on line-of-sight geometry, not absolute spatial distance. This is a well-tunneled sequence with a low DAS from both sides of the plate.

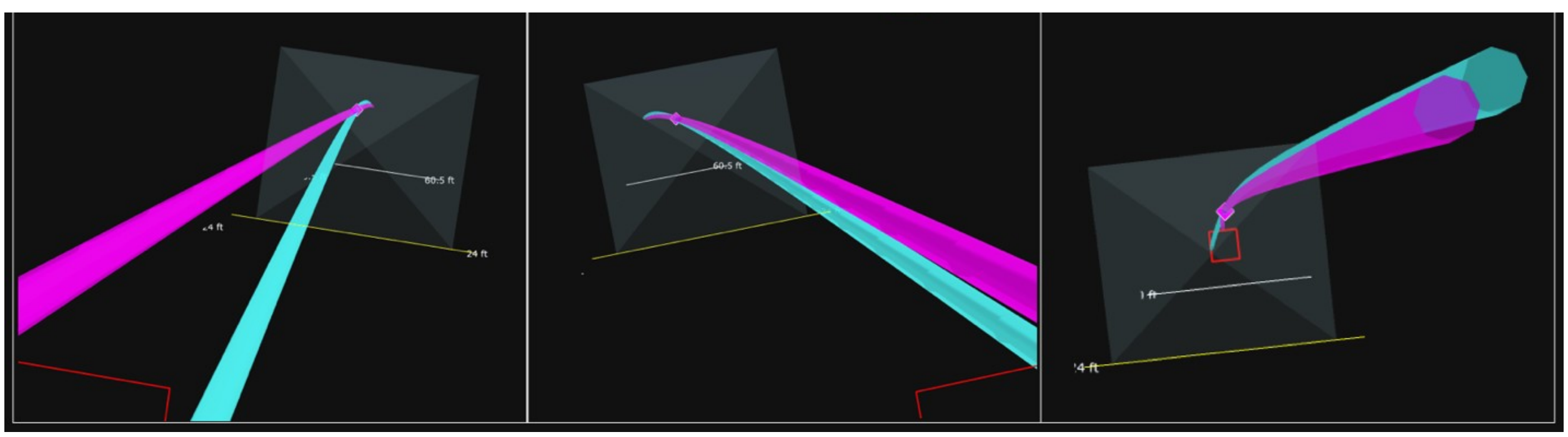

**Figure 2.** Two-pitch sequence from RHP Paul Skenes: left-handed hitter (panel 1) and right-handed hitter (panel 2).

Even small changes in vantage point alter the angular relationships between two pitch trajectories. What matters is not the physical distance between pitches in 3D space, but the visual angle they subtend at the batter's decision point. This motivates a batter-perspective geometry, which DAS/AAS compute directly. Though the effect of visual separation doesn't override the influence of the platoon advantage of hitting on the opposite side of the plate to the pitcher's throwing arm, it reiterates the idea of how much it matters to measure tunneling from the eyes of the batter, opposed measuring the pitch similarity in objective three-dimensional space. This visual dichotomy suggests that "deception" is not solely a physical property of the pitch (velocity/movement) but a geometric property of the matchup. A sequence that is deceptive to one batter may be easily decipherable to another simply due to the angle of observation.

This observation motivates our primary methodological shift, in that we cannot rely on raw coordinate differences ($\Delta x$, $\Delta z$) to measure deception. Instead, we must model the DAS from the batter's specific optical origin to accurately quantify the difficulty of the decision task.

## 2 Methods

Our analytical strategy combined inferential testing with predictive modeling. Bukiet, Harold, and Palacios (1997) modeled a half-inning as a Markov chain of base-out states. Stern (1991) modeled the probability of winning a football game from the score margin. Both treat a contest as a sequence of states rather than a set of independent plays. Our unit is shorter, two pitches inside one plate appearance, but the previous pitch still changes what the next pitch is worth.

The predictive models are gradient-boosted trees, fit as XGBoost, following Friedman (2001) and the standard machine-learning references in Bishop (2006) and James et al. (2021). A hierarchical Bayesian version of the same tunneling model would follow Gelman et al. (2013). That extension is left for later work.

Feature importance was assessed with SHAP values (Lundberg & Lee, 2017), so the contribution of DAS and AAS to a prediction can be read off the model directly.

### 2.1 Data Acquisition & Sample Construction

Our analysis uses Major League Baseball (MLB) Statcast Pitch-level tracking data from MLB Statcast records obtained from Baseball Savant's public Statcast Search CSV export (baseballsavant.mlb.com/statcast_search/csv) for games dated 2021-04-01 through 2025-08-09, accessed with the open-source Python client pybaseball.statcast().

We restricted the dataset to pitches thrown in standard game contexts and drop intentional walks, pitchouts, and rows with missing critical physics or location fields (release position, velocity/acceleration components, plate location, and batter-specific strike zone bounds). To exclude extremely low-volume or noncompetitive arms while retaining relievers and secondary pitches, we remove pitchers who simultaneously satisfy (i) fewer than 500 total pitches in the window and (ii) maximum observed fastball velocity below 86 mph. This filter is intentionally permissive relative to starter-only designs.

After chronological sorting by game, at-bat number, and pitch number, we construct two-pitch sequences within the same plate appearance by lagging pitch attributes one step. The analysis unit is therefore the ordered pair ($P_1$, $P_2$) thrown by the same pitcher to the same batter. We exclude (a) the first pitch of each plate appearance (no predecessor) and (b) pairs in which $P_1$ and $P_2$ share the same Statcast pitch classification (i.e. curveball–curveball pairs, slider–slider pairs are omitted from our dataset). The modeling sample therefore emphasizes cross-type sequencing, which is the setting where visual tunneling is strategically relevant. After all filters, the analytic file contains 740,936 two-pitch sequences.

A minimum threshold of 10 historical pitches per pitcher per zone was required to establish a valid baseline. While a higher threshold (e.g., $N \geq 50$) would produce a theoretically smoother, more continuous baseline distribution, it would severely bias the dataset by eliminating relievers, secondary pitches, and highly leveraged counts. We opted to retain these critical strategic situations. The resulting bimodal nature of the residual target variable mathematically constrains the maximum possible $R^2$, as the model predicts a continuous expectation against a discrete binary realization. Consequently, the predictive models are utilized primarily for relative feature validation (ablation), while the core magnitude of the deception effect is quantified via direct inferential testing and swing induction analysis.

For predictive models, we use an 80/20 random split within each sequence category (with a fixed seed for reproducibility). A random rather than temporal split is appropriate because the scientific target is a physics/perception relationship (angular geometry → batter response), not season-ahead forecasting. To mitigate the risk of 'identity leakage' (where a model might overfit to a specific pitcher's unique movement profile), we employed two safeguards:

1. The target variable, SwStr Lift, was calculated relative to a pitcher-context baseline (Section 2.3), effectively centering each pitcher's data around their own mean before the split.
2. Pitcher and batter IDs were excluded from the model's feature set.

This ensures that the model is learning the geometric properties of the sequence rather than memorizing individual player tendencies.

## 2.2 Feature Engineering

### *2.2.1 Pitch Group & Zone Taxonomy*

We derived several key predictive features from Statcast variables to characterize both pitch context and visual deception.

Based on the classified pitch type for each individual pitch in Statcast, pitches were bucketed into three categories:

| Group | Included Pitch Names |
|---|---|
| FB (Fastballs) | 4-Seam Fastball, Sinker, Cutter, Fastball |
| BR (Breaking) | Slider, Curveball, Sweeper, Slurve, Knuckle Curve |
| OS (Off-speed) | Changeup, Split-Finger, Forkball |

Location uses a **batter-scaled proportional attack-zone grid**. For each pitch we take that batter's top and bottom strike zone boundary values, define the strike zone midpoint and height, and assign integer zones using nested rectangles scaled to plate width (1.417 ft): Heart (zones 1–9), Shadow (11–19), Chase (21–29), and Waste (0). This matches the Heart / Shadow / Chase convention used in the swing-induction analyses. Granular zone-to-zone maps exclude Waste to reduce sparsity where deception is least decision-relevant.

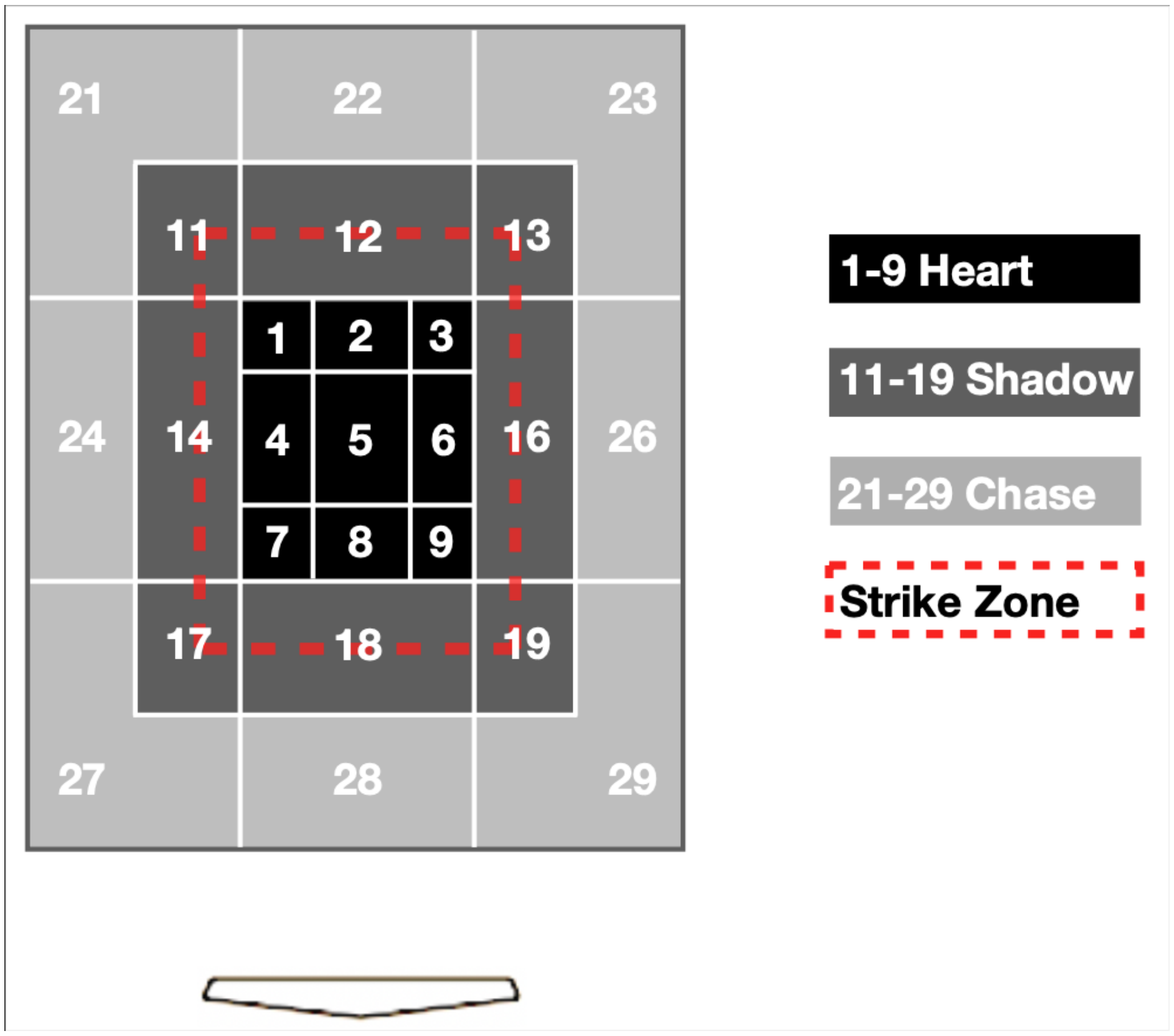

**Figure 3.** MLB Attack zone plot with the Heart, Shadow, Chase, and Waste zones. Waste was omitted for this project.

### *2.2.2 Trajectory reconstruction from Statcast kinematics*

Statcast reports kinematics referenced to a **50 ft** longitudinal plane, not the true release point. For every pitch we therefore:

1. Back-solve release kinematics. Using constant accelerations ($a_x$, $a_y$, a_z) and the reported 50 ft state ($v_{x0}$, $v_{y0}$, v_z0), solve the quadratic time offset between y = 50 and the observed release depth y, then adjust velocities to the true release.

    1. Integrate the full flight path. With release position and adjusted initial velocities, integrate the constant-acceleration equations from release to the front of home plate (y ≈ 0), sampling 50 points along the trajectory:

2. Interpolate at arbitrary y. Because angular metrics are defined on longitudinal stations, we interpolate (x, z) along each trajectory at target y values using monotone interpolation in y.

This pipeline converts raw Statcast kinematics into continuous 3-D paths suitable for batter-perspective geometry.

***2.2.3 Optical origin (batter eye position)***

Angular metrics are computed from a fixed optical origin E that depends on batter handedness (Statcast coordinates in feet):

• Right-handed batter (RHH): $E_R = (-1.5, -1.75, 4.5)$

• Left-handed batter (LHH): $E_L = (1.5, -1.75, 4.5)$

Sequences are processed separately by batter handedness so that DAS/AAS always use the matching eye position. Absolute Euclidean separation in space is not used as the deception measure; only the angle subtended at E.

***2.2.4 Decision Angular Separation (DAS) and Average Angular Separation (AAS)***

Our central predictive features are DAS and AAS. Both metrics are derived using a geometric framework in which the batter's eyes are treated as the origin point. Pitch trajectories were represented as vectors in three-dimensional space, and angular separation was computed as shown in Equation (1):

$$\theta(y) = \arccos( [v_1(y) \cdot v_2(y)] / [\|v_1(y)\| \, \|v_2(y)\|] ), \text{ where } v_i(y) = P_i(y) - E. \qquad (1)$$

Thus $P_1(y)$ and $P_2(y)$ denote the 3-D coordinates of pitches $P_1$ and $P_2$ evaluated at longitudinal position y (Statcast y decreases toward the plate). The batter's eye position E is treated as a fixed point in the same coordinate system (Section 2.2.3). We evaluate angular separation at the decision point $y_d = 24$ ft from home plate. Sayre et al. (2017) set the point of no return at 23.8 feet, the distance a league-average fastball travels in the last 175 ms before contact, and we round that depth to 24 feet. Thus $\theta(y_d)$ is the angle between the line-of-sight vectors at that decision depth. The cosine argument is clipped to [−1, 1] for numerical stability.

DAS is defined as the angular separation between $P_1$ and $P_2$ at the decision point, capturing how similar the two pitches appeared when the batter must commit to swing. AAS was defined as the average angular separation from the release point to the decision point, capturing how closely the pitches tracked together over the first portion of their flight. Gray (2017) is the reason for reporting both. Batters use optical information, including the angular expansion of the ball, continuously up to the decision point, so two pitches that meet at one depth can still have been distinguishable earlier. AAS summarizes that earlier evidence. DAS isolates the difference at commitment, about 175 ms before contact.

Formally, AAS averages $\theta(y)$ over the early flight path from $P_1$'s release depth y_rel to y_d, using 10 equally spaced stations in y:

$$\text{with } y_k = y_{rel} + ((k - 1)/9) \times (y_d - y_{rel}) \text{ for } k = 1, \ldots, 10.$$

### 2.3 Target Variable

Our outcome variable, Swinging Strike Lift (SwStr Lift), measures the marginal value of sequencing above pitch-level expectations. For each $P_2$, we observed whether it resulted in a swinging strike (coded 1 if yes, 0 otherwise). We then subtracted the baseline swinging-strike probability for that pitcher's pitch type, location, and count. This approach establishes a descriptive baseline of expected performance for that context, from which the lift is derived in Equation (2):

$$\text{SwStr Lift} = I(P_2 \text{ is swinging strike}) - \text{Baseline SwStr\%}. \tag{2}$$

Baselines were estimated using the conditional expectation of the full sample, stratified by Pitcher × Pitch Type × Zone × Balls × Strikes. A positive SwStr Lift indicates that the sequence produced more whiffs than expected, suggesting added value through deception. This design parallels variance decomposition methods (Kaplan, 2006) by separating intrinsic pitch quality from contextual sequencing effects.

Equivalently, for pitch $P_2$, let $I_{swstr} = 1$ if Statcast classifies the pitch as a swinging strike (including blocked swinging strikes), else 0. Define a context baseline as the empirical mean of $I_{swstr}$ over all pitches (not only sequenced pitches) in the cell

$$(\text{pitcher, pitch type, zone, balls, strikes}),$$

retaining cells with $N \geq 10$ pitches. Then

$$\text{SwStr Lift} = I_{swstr} - \hat{\Pr}(\text{swstr} \mid \text{cell}).$$

Subtracting the pitcher–pitch–zone–count mean removes the main effects of pitch quality and usage context so that remaining variation is attributable to *when* that pitch is thrown in a sequence (including tunneling geometry).

Because the baseline is estimated from all pitches in the cell, it includes both sequenced and non-sequenced offerings. Tunnel-driven whiffs therefore partially enter the baseline, which attenuates absolute lift toward zero (conservative bias) rather than inventing a false positive. Critically, within a fixed cell the baseline is constant, so contrasts between low-DAS and high-DAS sequences in the same context are equivalent to contrasts in raw swinging-strike rates. Complementary analyses that report raw Swing% and SwStr% by DAS/AAS quintile (Section 3.5) do not use this baseline at all.

### 2.4 Analytical Strategy

We evaluate hypotheses with four complementary layers:

1. Extreme-group tests (H1). Within each sequence category, label Good tunnels as the lowest 20% of DAS (or AAS) and Bad tunnels as the highest 20%. Compare mean SwStr Lift with Welch's t-tests (unequal variance).
2. Full quintile gradients (transparency). To ensure middle mass is not hidden by extreme-group designs, we also partition DAS/AAS into five equal-frequency quintiles (Q1 = tightest … Q5 = loosest) and report Swing% and SwStr% in All / Shadow / Chase regions (Figure 5).
3. Predictive ablation. For each major sequence category with over 200 sequences, train gradient-boosted tree regressors (100 trees, maximum depth 3, learning rate 0.1) predicting SwStr Lift with and without DAS/AAS; compare out-of-sample $R^2$.
4. Mechanism (H2). Contrast Swing% and SwStr% for Good vs Bad tunnels when $P_2$ lands in Shadow or Chase.

Layers (1) and (4) use the same Good/Bad tunnel definition (lowest vs highest 20% DAS/AAS) but answer different questions. Layer (1) tests H1 by comparing mean SwStr Lift across all landing locations. Layer (4) tests H2 by comparing raw Swing% and SwStr% only when $P_2$ lands in Shadow or Chase, which are the regions where swing induction is strategically relevant.

To show how predictions change as DAS or AAS vary while holding other features fixed, we compute Individual Conditional Expectation curves and their average (partial dependence) for the trained category models (Section 3.6). Exploratory zone-to-zone "best path" tables ranked on SwStr/Swing % Gap are prone to small-sample extremes. We therefore re-rank strategies with empirical-Bayes shrinkage of those Good-vs-Bad rate gaps. (Section 3.7).

## 3 Results

### 3.1 Descriptive Statistics

Before turning to hypothesis testing, we first summarize the basic behavior of the sequencing outcome, SwStr Lift. Because it is calculated by subtracting a fractional baseline expectation from a binary outcome (0 or 1), the resulting variable is mathematically constrained. Meaningful variation arises primarily from how much the realized swinging-strike outcome deviates from the highly specific pitch-context baseline. A cross-sequence summary reveals a modest positive central tendency (mean +0.8 percentage points), but substantial heterogeneity in lift across pitch-type transitions. This structural perspective also motivates the modeling considerations discussed later in Section 3.4: because SwStr Lift collapses onto two deterministic tracks, variance explained metrics such as $R^2$ must be interpreted relative to the inherent ceiling imposed by the target's construction.

### 3.2 Hypothesis Test Results: DAS and AAS

To directly test H1, we compared Good Tunnels (lowest 20% of values) and Bad Tunnels (highest 20%) for both DAS and AAS. Table 1 presents the mean SwStr Lift values for DAS, while Table 2 presents the same analysis for AAS. For both metrics, Good Tunnels significantly outperformed Bad Tunnels, with $p < 0.00001$ in nearly all categories.

For DAS, the FB→BR sequence yielded a +3.80 percentage-point difference in mean SwStr Lift; FB→OS produced +3.41 pp, FB→FB +1.46 pp, and OS→FB +1.23 pp. For AAS, effects were similar in direction and of comparable magnitude, with FB→BR slightly smaller (+3.45 pp vs +3.80 pp), confirming robustness of the tunneling signal.

**Table 1.** Comparison between Good tunnels (tightest 20% / lowest DAS) and Bad tunnels (loosest 20% / highest DAS) across pitch sequence categories.

| category | n_samples | Mean Lift (Good Tunnels) | Mean Lift (Bad Tunnels) | Lift Improvement | p_value |
|---|---|---|---|---|---|
| FB_OS | 102371 | 0.19% | -3.23% | 3.41% | < 0.0001 |
| FB_BR | 182351 | -0.06% | -3.86% | 3.80% | < 0.0001 |
| BR_OS | 42621 | 1.62% | -0.63% | 2.25% | < 0.0001 |
| BR_BR | 23858 | 2.42% | 0.32% | 2.10% | 0.00550 |
| OS_BR | 23365 | -0.08% | -2.06% | 1.98% | 0.00552 |
| FB_FB | 99331 | 0.62% | -0.84% | 1.46% | < 0.0001 |
| OS_FB | 81164 | -0.02% | -1.25% | 1.23% | < 0.0001 |
| BR_FB | 185761 | 0.23% | -0.36% | 0.58% | 0.00415 |

*Note.* Good tunnels consistently yielded higher SwStr Lift, confirming H1.

**Table 2.** T-test validation of AAS.

| category | n_samples | Mean Lift (Good Tunnels) | Mean Lift (Bad Tunnels) | Δ Mean Lift % | p_value |
|---|---|---|---|---|---|
| FB_OS | 102371 | 0.26% | -3.28% | 3.55% | < 0.0001 |
| FB_BR | 182351 | -0.18% | -3.63% | 3.45% | < 0.0001 |
| BR_OS | 42621 | 1.80% | -0.52% | 2.32% | < 0.0001 |
| OS_BR | 23365 | 0.18% | -1.98% | 2.16% | 0.00246 |
| BR_BR | 23858 | 2.20% | 0.38% | 1.81% | 0.01630 |

| category | n_samples | Mean Lift (Good Tunnels) | Mean Lift (Bad Tunnels) | Δ Mean Lift % | p_value |
|---|---|---|---|---|---|
| FB_FB | 99331 | 0.53% | -0.88% | 1.41% | < 0.0001 |
| OS_FB | 81164 | -0.01% | -1.32% | 1.32% | < 0.0001 |
| BR_FB | 185761 | -0.17% | -0.41% | 0.24% | 0.23510 |

Figures 4.1 and 4.2 provide visual corroboration by showing higher SwStr% and Swing% for tighter tunnels in the Shadow and Chase regions. The full-quintile gradients in Figure 5 provide an additional check that these differences are not confined to the extreme groups.

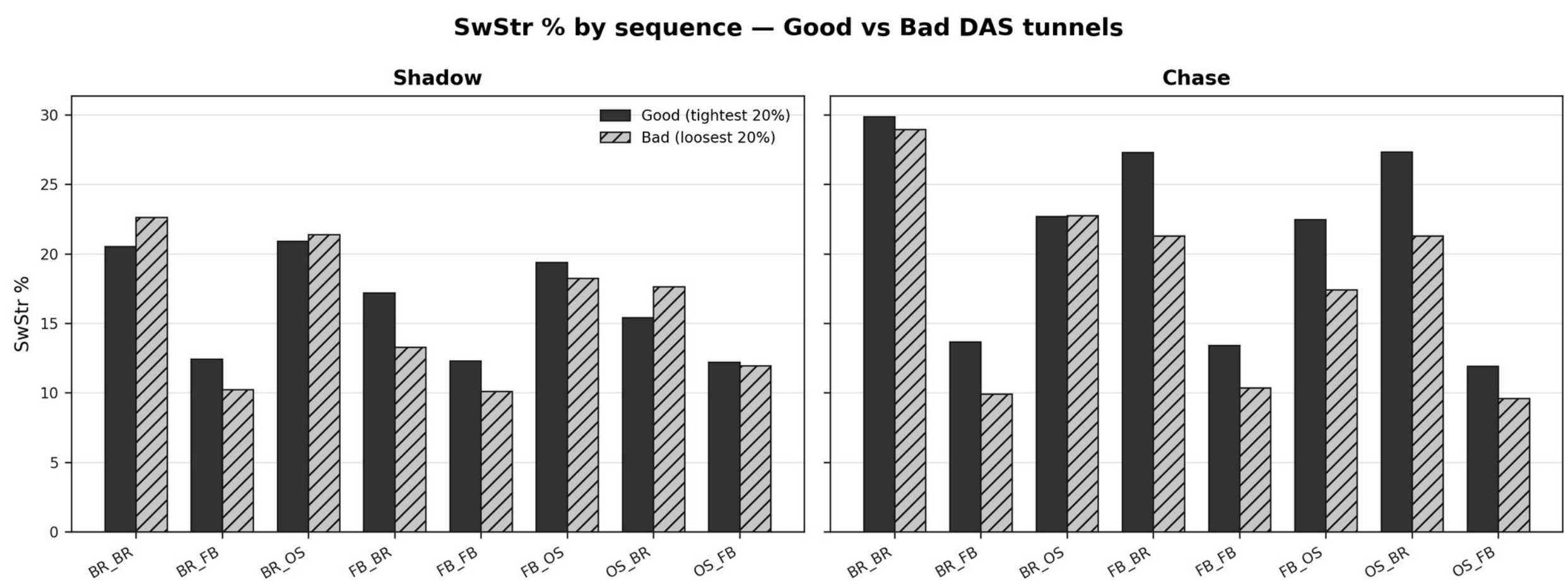

**Figure 4.1.** Bar charts of swinging-strike rate (SwStr%) in the Shadow and Chase zones for Good vs Bad DAS tunnels, by pitch-sequence category.

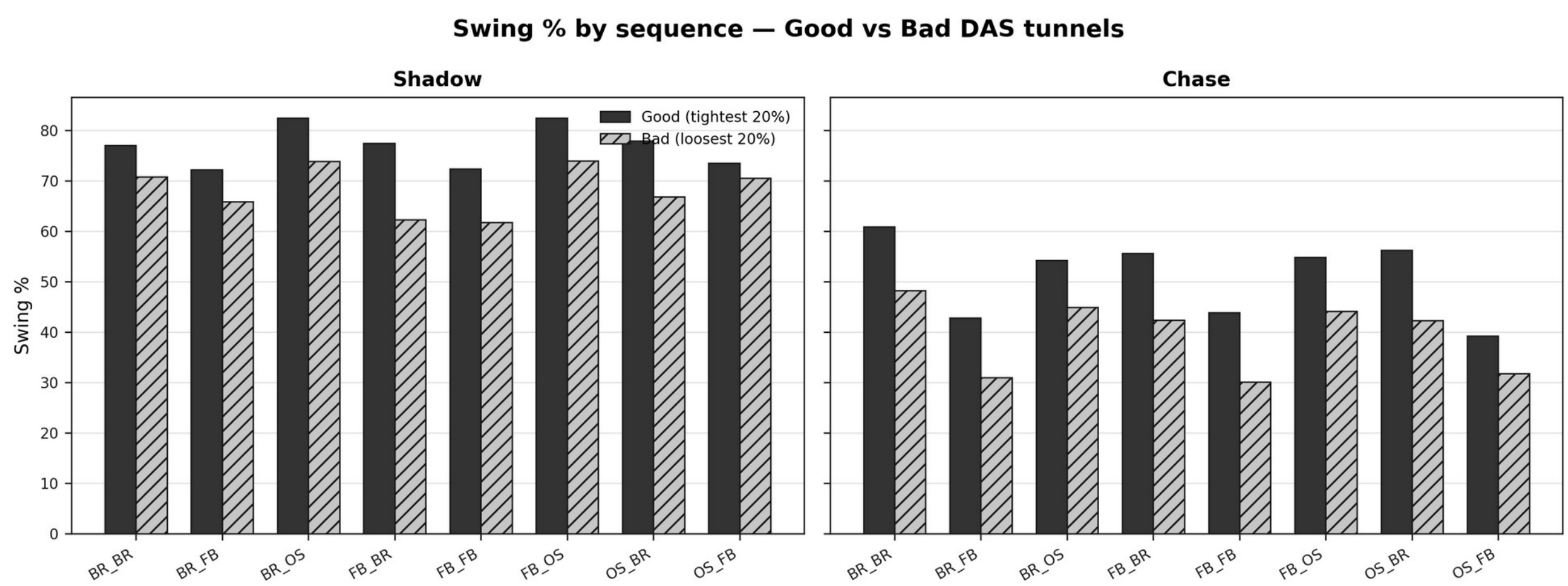

**Figure 4.2.** Bar charts of swing rate (Swing%) in the Shadow and Chase zones for Good vs Bad DAS tunnels, by pitch-sequence category.

### 3.3 Swing Induction as the Primary Behavioral Pathway of Tunneling

A central implication of the tunneling hypothesis is that geometric similarity between consecutive pitches should not simply improve whiff probability conditional on swinging, but that it should change the batter's underlying swing behavior. When two pitches remain visually indistinguishable at the decision point, the hitter faces reduced precommit information and is likely to expand their actionable swing zone. This *swing induction* mechanism provides the pathway through which deception translates into measurable sequencing value.

Our empirical results strongly support this channel. For the common FB→BR use case, well-tunneled sequences with $P_2$ in Shadow or Chase (and $P_1$ outside Waste) produced a +22.3 percentage-point higher swing rate than poorly tunneled sequences (67.5% vs. 45.2%).

This effect is not only statistically overwhelming but strategically consequential: hitters are ordinarily highly selective in these zones due to the lower expected run values in these regions. Inducing additional swings at pitches outside the Heart zone directly decreases the expected run value of an outcome for the hitter, whether that be a swinging strike or weak contact.

The pattern generalizes across other pitch-type combinations, but the magnitude varies with pitch shapes and sequencing roles. Importantly, swing induction aligns precisely with the geometric principles introduced in Sections 1.3 and 1.4. Low-DAS and low-AAS sequences reduce angular separation at the decision point, compressing the batter's discriminability of pitch type and trajectory. This perceptual bottleneck forces earlier or less-informed swing commitments, producing systematic increases in swing rate in precisely the regions (Shadow/Chase) where hitters typically avoid offering.

**Table 3.** SwStr% Gap by zone.

| Sequence | N_Good | N_Bad | SwStr% Good | SwStr% Bad | SwStr% Gap |
|---|---|---|---|---|---|
| FB→BR | 25,588 | 7,513 | 22.1% | 13.6% | +8.5% |
| OS→BR | 3,122 | 754 | 21.3% | 14.2% | +7.1% |
| BR→FB | 23,061 | 12,162 | 13.7% | 7.7% | +6.0% |
| FB→OS | 17,103 | 4,434 | 20.7% | 16.3% | +4.5% |
| OS→FB | 11,179 | 3,630 | 12.4% | 9.3% | +3.2% |
| FB→FB | 14,699 | 6,937 | 12.9% | 9.8% | +3.1% |
| BR→OS | 5,735 | 2,750 | 20.7% | 23.4% | -2.6% |
| BR→BR | 2,778 | 1,148 | 23.4% | 28.2% | -4.8% |

*Note.* Tightest 20% = lowest DAS/AAS (Good); loosest 20% = highest DAS/AAS (Bad).

**Table 4.** Swing% Gap by zone.

| Sequence | N_Good | N_Bad | Swing% Good | Swing% Bad | Swing% Gap |
|---|---|---|---|---|---|
| OS→FB | 11,179 | 3,630 | 65.7% | 36.3% | +29.4% |
| FB→FB | 14,699 | 6,937 | 65.0% | 38.7% | +26.3% |
| OS→BR | 3,122 | 754 | 68.4% | 43.2% | +25.1% |
| FB→OS | 17,103 | 4,434 | 71.2% | 46.1% | +25.1% |
| BR→FB | 23,061 | 12,162 | 66.7% | 42.6% | +24.1% |
| FB→BR | 25,588 | 7,513 | 67.5% | 45.2% | +22.3% |
| BR→OS | 5,735 | 2,750 | 71.0% | 52.8% | +18.2% |
| BR→BR | 2,778 | 1,148 | 70.5% | 56.4% | +14.0% |

Taken together, these results establish swing induction as the core behavioral mechanism by which tunneling generates sequencing value. Subsequent sections (3.4 and 4.2) show how this mechanism helps explain both the predictive importance of DAS and its strategic role in modern pitch design and game planning.

### 3.4 Predictive Modeling and Feature Importance

We next assessed predictive value through ablation studies using gradient-boosted tree models. Because SwStr Lift is defined as a residualized binary outcome, each observation is mathematically restricted to one of two deterministic tracks: (1|Baseline) for a swinging strike, and (0|Baseline) otherwise. This produces a highly constrained variance structure. Unlike continuous-valued regression targets where predictions can neatly map to observations, our models are predicting a continuous probability against a discrete, bimodal realization. As a result, $R^2$ cannot be interpreted in the traditional sense; the inherent variance of a Bernoulli-derived residual imposes a strict mathematical ceiling on the maximum possible $R^2$. Accordingly, the more appropriate indicators of model performance in this context are relative $R^2$ improvements between baseline and augmented models.

Baseline models included conventional predictors (pitch pair differentials, current count, previous pitch zone). Augmented models included DAS and AAS. When evaluating the binary classification performance of the underlying probability model, the inclusion of DAS and AAS confirmed that the geometric features provide information gain that improves the calibration of swinging strike probabilities, not just the ranking of sequences. As shown in Tables 5a and 5b, the inclusion of tunneling metrics substantially improved out-of-sample $R^2$ for fastball-initiated sequences (FB→BR, FB→OS), but yielded negligible or slightly negative changes in sequences initiated by breaking balls (BR→BR, BR→FB).

For FB→BR sequences, $R^2$ improved from 0.0040 to 0.0056 (+38.2%). For FB→OS sequences, $R^2$ improved from 0.0019 to 0.0049 (+162.8%). Though absolute $R^2$ values remain small, these relative gains are meaningful and align with expectations for highly stochastic pitch outcomes.

**Table 5a.** Ablation study results with tunneling features (DAS/AAS) included.
With Tunneling:

| category | n_samples | R2 | MSE | RMSE |
|---|---|---|---|---|
| FB_OS | 102371 | 0.004881 | 0.117286 | 0.342470 |
| OS_FB | 81164 | 0.000704 | 0.079759 | 0.282416 |
| FB_BR | 182351 | 0.005582 | 0.124755 | 0.353206 |
| BR_FB | 185761 | -0.000530 | 0.079544 | 0.282036 |
| BR_BR | 23858 | -0.002710 | 0.138410 | 0.372034 |
| OS_BR | 23365 | 0.001167 | 0.124806 | 0.353280 |
| FB_FB | 99331 | 0.000796 | 0.082595 | 0.287393 |
| BR_OS | 42621 | -0.000933 | 0.126906 | 0.356238 |

**Table 5b.** Ablation study results without tunneling features (baseline predictors only).
Without Tunneling

| category | n_samples | R2 | MSE | RMSE |
|---|---|---|---|---|
| FB_OS | 102371 | 0.001857 | 0.114069 | 0.337741 |
| OS_FB | 81164 | -0.000258 | 0.080419 | 0.283582 |
| FB_BR | 182351 | 0.004040 | 0.126373 | 0.355490 |
| BR_FB | 185761 | 0.000111 | 0.079996 | 0.282836 |
| BR_BR | 23858 | -0.001002 | 0.145014 | 0.380808 |
| OS_BR | 23365 | -0.006229 | 0.134502 | 0.366745 |
| FB_FB | 99331 | 0.000216 | 0.082113 | 0.286554 |
| BR_OS | 42621 | -0.000776 | 0.126981 | 0.356343 |

*Note.* A ~1% variation in sample size between models reflects rows dropped for missing predictors. See Tables 5a and 5b.

Feature importance analysis reinforced these results. SHAP plots (Figure 7) demonstrated that DAS consistently ranked among the top predictors, often rivaling velocity and pitch location.

Low DAS values were associated with positive SHAP contributions, confirming their predictive utility.

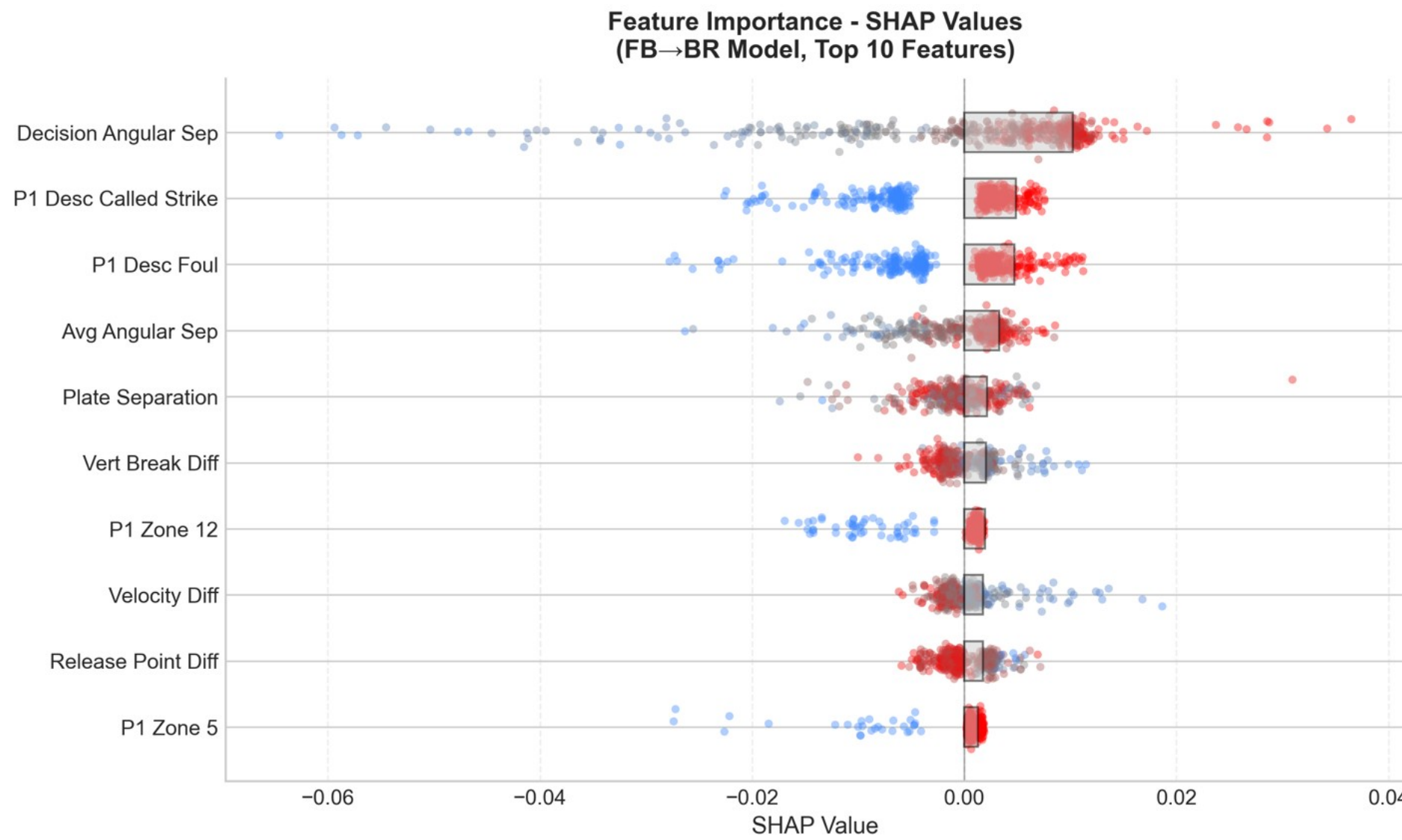

**Figure 7.** Bootstrapped SHAP feature importance for FB→BR sequences, highlighting DAS/AAS among the strongest predictors of SwStr Lift relative to conventional pitch features.

### 3.5 Full Quintile Gradients

Extreme-group tests maximize contrast but can hide non-monotonicity in the middle of the distribution. We therefore also plot Swing% and SwStr% across DAS (and AAS) quintiles Q1 (tightest 20%) through Q5 (loosest 20%) for major sequence categories, including All / Shadow / Chase strata.

For the strategically central FB→BR Chase stratum under DAS, rates decline from Q1 to Q5 rather than appearing only at the tails: SwStr% moves from 27.3% (Q1) through a flat mid-band (~27% in Q2–Q3) to 25.3% (Q4) and 21.3% (Q5); Swing% falls more smoothly from 55.6% to 42.4%. This pattern indicates that (i) the Good-vs-Bad signal is not an artifact of discarding the middle 60%, and (ii) much of the swinging-strike separation accumulates in the poorest tunnels (Q4–Q5), while swing induction deteriorates more continuously as visual separation increases.

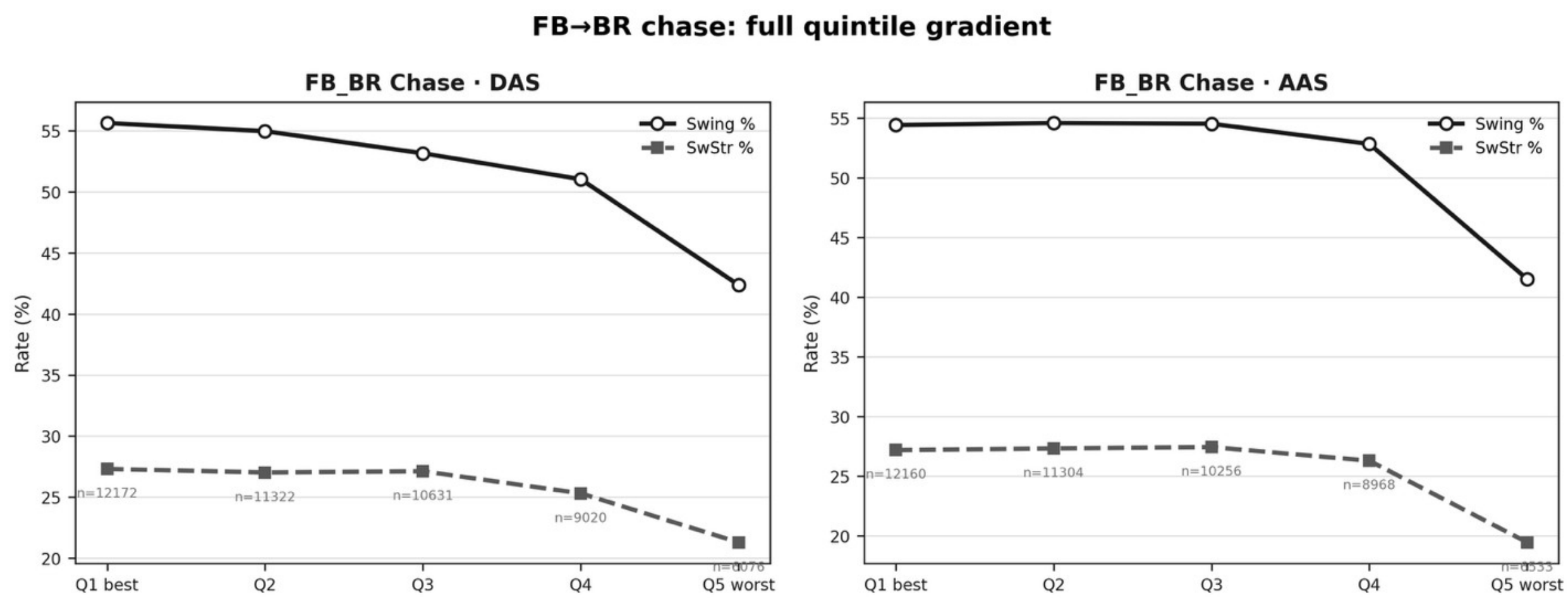

**Figure 5.** Full quintile gradients of SwStr Lift across DAS/AAS bins for key sequence strata (including FB→BR Chase), showing that tunneling gains rise continuously rather than only at the extremes.

### 3.6 How Model Predictions Respond to DAS and AAS

SHAP establishes importance; ICE establishes shape. Individual Conditional Expectation curves and partial dependence for the FB→BR model show how predictions change when DAS or AAS is swept with all other features held fixed. For DAS, partial dependence falls from roughly +2.3 percentage points predicted SwStr Lift at tight tunnels (~0.2°) to about −7.0 percentage points at wide separation (~5°), an ≈9.4 percentage-point swing in model expectation. AAS shows a milder negative slope (~2 percentage points over its typical range). Centered ICE plots (change relative to each sequence's observed value) confirm that moving toward tighter tunnels raises predicted lift for most observations. These response surfaces operationalize H1 inside the predictive model and address the practical question: if DAS/AAS change, how does the prediction change?

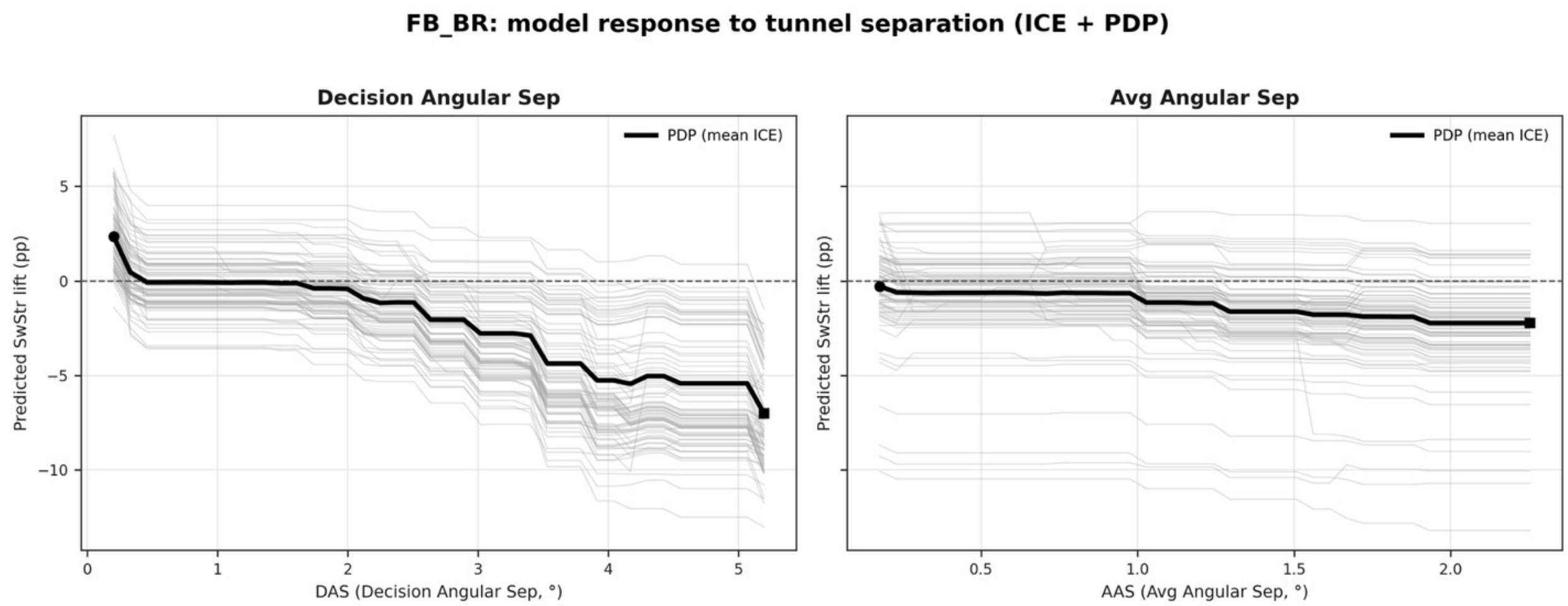

**Figure 6.** Individual conditional expectation (ICE) curves and partial dependence of predicted SwStr Lift on DAS and AAS, illustrating how model predictions respond as visual separation changes.

### 3.7 Reliability-Weighted Zone-to-Zone Strategies

Unweighted rankings of granular zone paths by raw SwStr% Gap preferentially surface low-frequency cells (winner's curse): extreme Good−Bad contrasts that are statistically fragile. Figure 9 addresses this with empirical-Bayes (EB) shrinkage of those Gaps toward the null, down-weighting paths with high sampling variance.

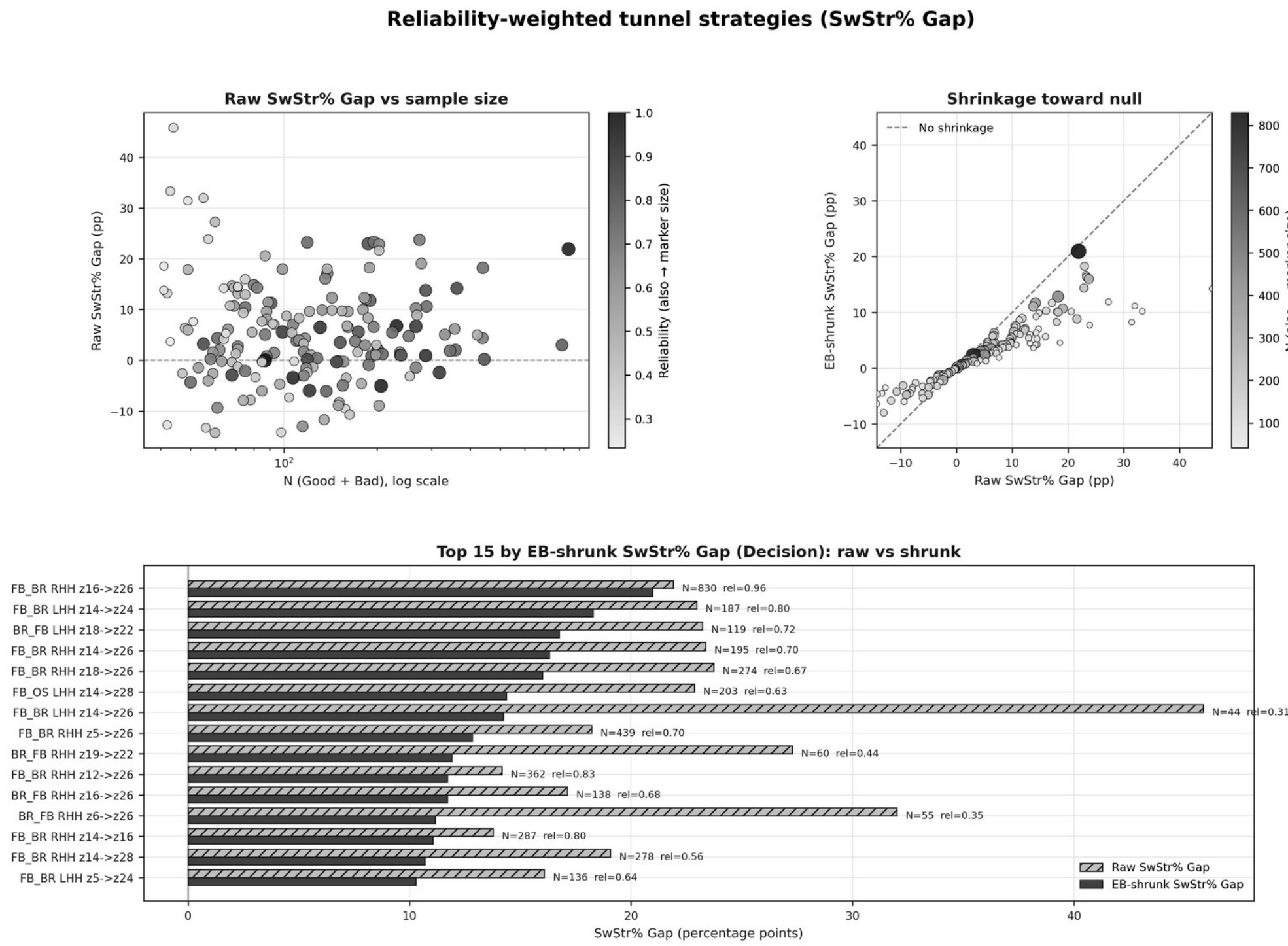

**Figure 9.** Reliability-weighted (empirical-Bayes shrunk) zone-to-zone SwStr% Gap rankings, down-weighting high-variance paths relative to raw Gaps.

The top-left panel shows raw SwStr% Gap against sample size *N* (Good + Bad), with marker size/shade indicating reliability. The familiar funnel appears: small-*N* paths fan out to large raw Gaps, while high-*N* paths concentrate at more moderate values. The top-right panel plots EB-shrunk Gap against raw Gap; points fall toward zero relative to the no-shrinkage diagonal, with low-*N* cells pulled hardest. The bottom panel ranks the top paths by shrunk Gap and contrasts raw (hatched) versus shrunk (solid) estimates, annotated with *N* and reliability.

After shrinkage, top-ranked strategies become more sample-supported and less extreme (e.g., top-25 mean SwStr% Gap ~23 pp raw → ~12 pp shrunk in the decision-metric granular set under our filters). High-reliability routes such as FB→BR to RHH, Shadow→Chase (zone 16→26; $N \approx$ 830, reliability ≈ 0.96) retain large Gaps near their raw values, whereas several raw "+40–100%" paths are heavily attenuated or drop out of the ranking. We therefore treat EB-shrunk rankings as the preferred exploratory guide for pitch design.

### 3.8 Exploratory Analysis: Top Sequences and Zone Combinations

To further contextualize tunneling's practical implications, we examined the top-performing pitch/zone/handedness combinations by SwStr% Gap. Tables 6 and 7 report the ten most effective combinations for DAS and AAS, respectively. These exploratory analyses highlight tactical patterns; for example, fastballs in the Shadow zones followed by well-tunneled breaking balls in the chase zone consistently generated a stronger SwStr%, regardless of batter handedness.

**Table 6.** Top 10 SwStr combinations by DAS.

| Sequence | N (Total) | Batter Hand | p1_zone | p2_zone | SwStr% (Bad) | SwStr% (Good) | SwStr% Gap |
|---|---|---|---|---|---|---|---|
| FB_BR | 2,619 | LHH | Shadow | Chase | 8.8% | 30.6% | +21.8% |
| FB_BR | 4,833 | RHH | Shadow | Chase | 10.9% | 26.3% | +15.4% |
| BR_FB | 2,646 | RHH | Heart | Shadow | 4.5% | 17.8% | +13.3% |
| BR_FB | 1,281 | LHH | Heart | Chase | 4.4% | 16.0% | +11.7% |
| FB_OS | 2,623 | RHH | Shadow | Chase | 13.2% | 24.2% | +11.0% |
| FB_OS | 2,360 | LHH | Shadow | Chase | 11.0% | 21.3% | +10.2% |
| BR_FB | 2,048 | RHH | Heart | Chase | 5.7% | 15.8% | +10.1% |
| BR_FB | 3,025 | RHH | Shadow | Chase | 7.4% | 17.3% | +9.9% |
| FB_BR | 4,669 | RHH | Shadow | Shadow | 8.8% | 18.2% | +9.4% |
| BR_FB | 2,028 | LHH | Shadow | Chase | 6.1% | 15.0% | +8.9% |

**Table 7.** Top 10 SwStr combinations by AAS.

| Sequence | N (Total) | Batter Hand | p1_zone | p2_zone | SwStr% (Bad) | SwStr% (Good) | SwStr% Gap |
|---|---|---|---|---|---|---|---|
| FB_BR | 2,027 | LHH | Heart | Chase | 6.3% | 29.7% | +23.4% |
| FB_BR | 2,864 | LHH | Shadow | Chase | 7.7% | 30.7% | +23.0% |
| FB_BR | 3,335 | RHH | Heart | Chase | 8.2% | 26.7% | +18.6% |
| BR_FB | 1,558 | LHH | Heart | Shadow | 5.6% | 18.6% | +13.0% |
| FB_OS | 2,695 | RHH | Shadow | Chase | 12.5% | 24.6% | +12.1% |
| FB_BR | 5,084 | RHH | Shadow | Chase | 14.3% | 26.3% | +12.0% |
| BR_FB | 2,481 | RHH | Heart | Shadow | 7.4% | 17.6% | +10.2% |
| FB_OS | 2,402 | LHH | Shadow | Chase | 10.9% | 20.9% | +10.0% |
| FB_FB | 1,103 | RHH | Heart | Chase | 6.4% | 16.1% | +9.7% |
| FB_BR | 3,242 | RHH | Heart | Shadow | 9.6% | 19.1% | +9.4% |

While exploratory, these results illustrate how tunneling metrics can be used not only for evaluation but also to inform game strategy and individualized pitch design.

To translate these findings into actionable game strategy, we mapped two-pitch zone transitions on the strike-zone grid (Figure 8). Circular markers denote the first-pitch ($P_1$) zone and squares the second-pitch ($P_2$) zone; marker size and path thickness scale with on-path pitch count. Paths through the waste zone are excluded. Within each matchup, tunnels are defined by the lowest versus highest 20% of DAS. Optimal paths further require min(NGood ,NBad ) ≥ 50, a positive Gap, and are limited to the top-ranked paths under that metric.

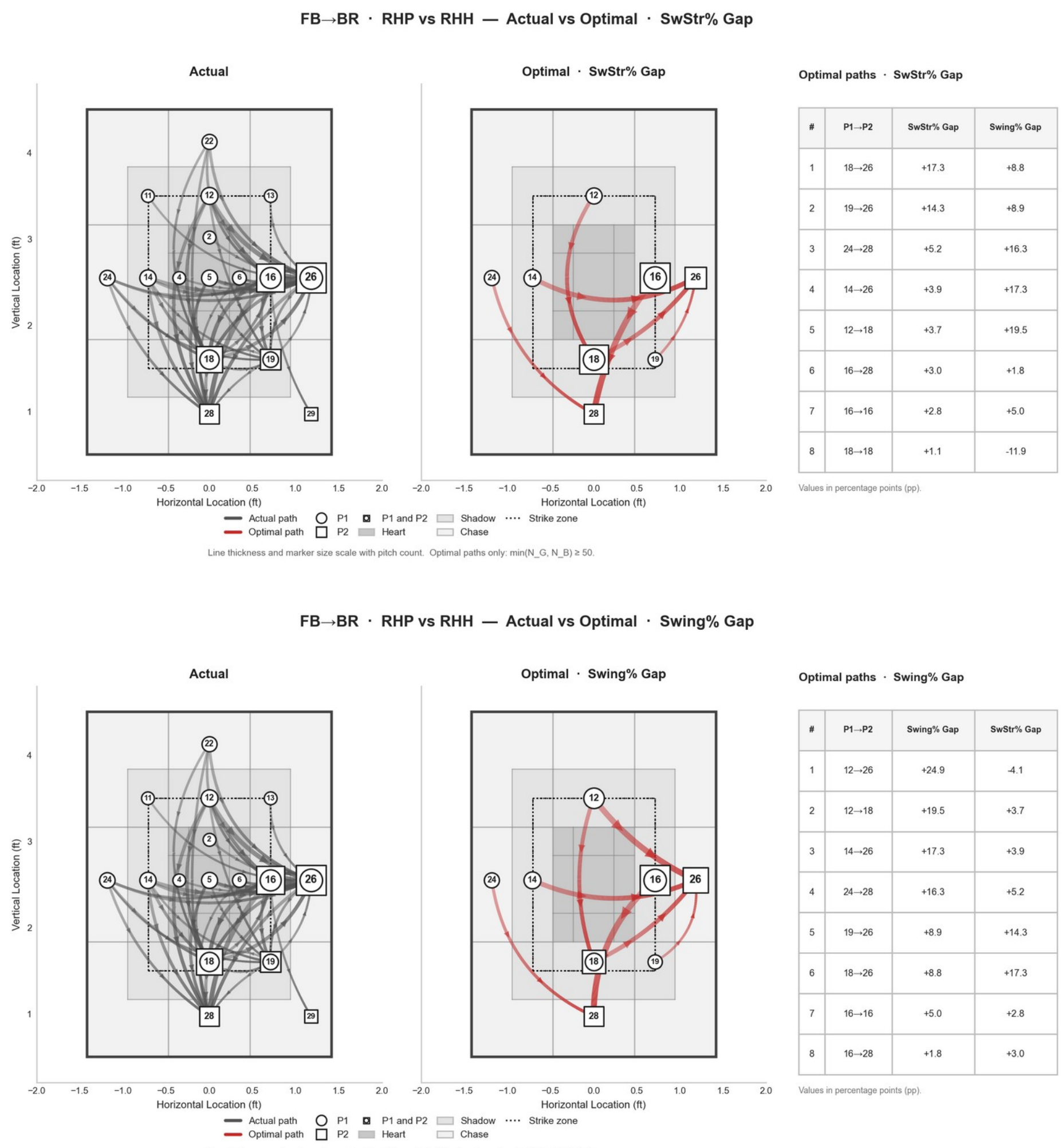

| # | P1→P2 | SwStr% Gap | Swing% Gap |
|---|---|---|---|
| 1 | 18→26 | +17.3 | +8.8 |
| 2 | 19→26 | +14.3 | +8.9 |
| 3 | 24→28 | +5.2 | +16.3 |
| 4 | 14→26 | +3.9 | +17.3 |
| 5 | 12→18 | +3.7 | +19.5 |
| 6 | 16→28 | +3.0 | +1.8 |
| 7 | 16→16 | +2.8 | +5.0 |
| 8 | 18→18 | +1.1 | -11.9 |

| # | P1→P2 | Swing% Gap | SwStr% Gap |
|---|---|---|---|
| 1 | 12→26 | +24.9 | -4.1 |
| 2 | 12→18 | +19.5 | +3.7 |
| 3 | 14→26 | +17.3 | +3.9 |
| 4 | 24→28 | +16.3 | +5.2 |
| 5 | 19→26 | +8.9 | +14.3 |
| 6 | 18→26 | +8.8 | +17.3 |
| 7 | 16→16 | +5.0 | +2.8 |
| 8 | 16→28 | +1.8 | +3.0 |

**Figure 8.** Top individual zone-to-zone tunneling impacts: Actual vs Optimal targeting panels summarizing where good tunnels most increase SwStr% Gap.

The left panel shows Actual Targeting: observed sequences weighted by frequency. Pitchers disperse $P_1 \rightarrow P_2$ paths widely across Heart, Shadow, and Chase.

The right panel shows Optimal Targeting for the panel metric (SwStr% Gap or Swing% Gap): Good-Bad rate contrasts for the same zone path, not residual SwStr Lift. All optimal paths are drawn in the same ink; Gap magnitude is reported in the adjacent ranked table rather than by color. Relative to the Actual panel, this filter isolates a smaller set of high-leverage routes. For the common FB→BR, RHP-vs-RHH case under SwStr% Gap, the leading paths disproportionately start toward the Heart or inner Shadow and finish in the low Shadow and Chase (e.g., 18→26, 19→26). That geometry is consistent with swing induction (H2): establish the zone centrally, then use a tight tunnel to force a commitment on a pitch that exits below the strike zone.

## 4 Discussion

Our findings validate that tunneling provides a measurable sequencing advantage. The broader relevance of this approach parallels strategic optimization studies in football (Romer, 2006), which demonstrate that subtle inefficiencies can be quantified and exploited in competitive contexts.

Park factors, pitch framing, and baserunning are all effects that a raw total can hide. Tunneling belongs in that group.

### 4.1 Evidence of the Tunneling Effect

The inferential results provide direct statistical validation that tunneling creates a measurable sequencing advantage. Welch's t-tests comparing the top 20% of sequences by tunnel quality (low DAS) against the bottom 20% (high DAS) revealed consistent and statistically significant differences ($p < 0.00001$ across most categories).

As shown in Table 1, for the common fastball-to-breaking ball (FB→BR) sequence, well-tunneled pairs generated a +3.80% improvement in mean SwStr Lift compared to poorly tunneled pairs. Other common sequences showed gains of +3.41% (FB→OS), +1.46% (FB→FB), and +1.23% (OS→FB). Though these values may appear modest in isolation, across thousands of pitches per season they aggregate into a practically meaningful advantage in pitcher effectiveness.

Visual corroboration strengthens these findings. Distributional displays of SwStr Lift are shifted upward for Good Tunnels, confirming the effect extends beyond averages to the entire distribution. Complementary scatter displays illustrate a strong positive correlation between the raw swinging-strike rate and SwStr Lift under low-DAS and low-AAS conditions (Figure 4.1, 4.2)

Importantly, robustness checks using Average Angular Separation (AAS) yielded similar results (Table 2). While effect sizes were slightly attenuated relative to DAS, the directional consistency across both measures reinforces confidence that tunneling effects are not an artifact of a single metric but reflect a generalizable underlying phenomenon.

### 4.2 Swing Induction as the Core Mechanism

While Section 3.3 establishes swing induction as the behavioral pathway through which tunneling exerts measurable influence, its broader implications extend beyond individual sequences. The mechanism reshapes the pitcher–batter information game: by reducing early-flight discriminability, tunneling increases hitter uncertainty, forces earlier commitment, and amplifies the value of well-designed sequencing strategies. This positions tunneling not merely as a micro-level effect but as a structural tool that interacts with pitch design, repertoire optimization, and count-dependent game theory.

This mechanism aligns with broader sabermetric work on hidden contributions. Baumer (2009) found that baserunning changes run scoring by less than ±25 runs in a typical season, an effect that is easy to miss and still worth measuring. Tunneling is the pitching version of that idea. Zone-level breakdowns (Tables 3 and 4) show that the effect is largest where the batter has the most discretion, on the margins of the strike zone.

Exploratory analyses extend these findings by identifying tactical patterns that consistently exploit swing induction. For example, Top-10 sequence combinations (Tables 6 and 7) revealed that right-handed pitchers pairing elevated fastballs with sliders down-and-away to same-handed hitters repeatedly ranked among the most effective deception-inducing patterns. This illustrates how tunneling metrics not only describe a general phenomenon but also provide concrete strategic insights.

### 4.3 Quantifying the Predictive Power of Tunneling

The predictive modeling results further establish tunneling as a unique and quantifiable skill. Ablation studies demonstrated that including DAS and AAS in gradient-boosted tree models substantially improved out-of-sample fit. For FB→BR sequences, $R^2$ increased from 0.0040 to 0.0056 (+38.2%), while for FB→OS sequences, $R^2$ improved from 0.0019 to 0.0049 (+162.8%). While the absolute values remain modest due to the inherent stochasticity of pitch outcomes, the relative improvements highlight the explanatory power of tunneling features.

Ablation studies (§3.4; Tables 5a and 5b) show that including DAS and AAS in gradient-boosted tree models substantially improves out-of-sample fit for fastball-initiated sequences, with the largest relative gains for FB→BR and FB→OS. While the absolute $R^2$ values remain modest due to the inherent stochasticity of pitch outcomes, the relative improvements highlight the explanatory power of tunneling features. SHAP summaries (Figure 7) reinforce that low DAS contributes positively to predicted SwStr Lift, often alongside velocity and location. Exploratory zone/handedness rankings further show how these model insights map onto actionable sequencing patterns.

SHAP feature importance plots (Figure 7) reinforce these findings by showing that low DAS values (blue points) consistently correspond with positive SHAP contributions. Well-tunneled sequences increase predicted SwStr lift, often rising or exceeding the predictive influence of traditional metrics such as velocity and pitch location. This elevates tunneling from an intuitive coaching heuristic to a measurable, model-relevant feature.

Moreover, the Top-10 exploratory analyses demonstrate how predictive insights can translate into applied contexts. By isolating pitch/zone/handedness combinations most associated with deception, tunneling metrics serve dual roles: improving model fit and providing directly actionable knowledge for game strategy and player development.

### 4.4 Limitations and Future Directions

Despite robust results, several limitations warrant discussion. Firstly, our framework is intentionally parsimonious, adopting an amnesiac assumption that only the preceding pitch affects the next outcome. In reality, an aspect to the art of pitching is determining ways to get batters out across three or more pitches based on how the pitcher feels, how lineups are reacting, and earlier results. Future work could relax this assumption using Markov chain or Bayesian sequence models.

Second, batter-level heterogeneity was not modeled. Hitters may systematically differ in their susceptibility to tunneling, and hierarchical models could illuminate these patterns. Third, reliance on Statcast data introduces the possibility of measurement error, which could bias DAS and AAS calculations. Fourth, SwStr Lift baselines are estimated from all pitches in a pitcher–pitch–zone–count cell, including sequenced pitches. This induces mild attenuation of absolute lift if tunneling itself elevates cell means; within-cell Good-vs-Bad contrasts remain well-defined because the baseline cancels. Raw quintile gradients (Section 3.5) provide a baseline-free corroboration. Fifth, ICE varies one angular metric while holding the other fixed, whereas DAS and AAS co-move in real sequences; ICE answers a model-sensitivity question, not a claim of independent causal levers. Finally, the outcome variable focuses exclusively on swinging strikes. While useful for isolating sequencing effects, tunneling may also influence other desirable outcomes correlating to run prevention, particularly inducing weak contact. Broader outcome modeling is needed.

Future research should therefore extend beyond whiffs to capture the full spectrum of pitch outcomes, incorporate batter-specific responses, and explore tunneling effects under different contextual conditions such as count leverage, pitcher fatigue, and game state. These extensions would align with the applied quantitative baseball research tradition of developing context-sensitive, rigorously quantified analytic tools to uncover hidden dimensions of performance.

## 5 Conclusion

This study formalizes DAS and AAS as batterperspective geometric measures of visual pitch similarity and introduces SwStr Lift to quantify the sequencing value beyond pitch level expectations. The proposed behavioral pathway is simple and testable: tight tunnels compress pre-commit information and are associated with greater swing propensity in the regions where hitters are most selective. Predictive models independently confirm this mechanism: DAS and AAS consistently improve outofsample accuracy and emerge as top predictors of sequencing value. These findings demonstrate that tunneling is not merely a descriptive artifact or coaching intuition, but a quantifiable skill that can be measured, monitored, and operationalized.

Our results position tunneling as a measurable, strategically actionable dimension of modern pitching, and highlight perceptual geometry as a promising frontier for future baseball analytics. More broadly, the framework extends naturally to pitch design, repertoire optimization, and sequence planning. By situating tunneling within the batter's perceptual constraints, we connect traditional sabermetric insights with modern trajectory-based analytics, showing that deception emerges not only from pitch quality but from the visual continuity between consecutive offerings. Transparency analyses (full quintile gradients, ICE response curves, and reliability-weighted strategy maps) reinforce that tunneling effects are graded, model-visible, and not an artifact of extreme-group or small-sample ranking designs.

## References

Acharya, R. A., Ahmed, A. J., D'Amour, A. N., Lu, H., Morris, C. N., Oglevee, B. D., Peterson, A. W., & Swift, R. N. (2008). Improving Major League Baseball park factor estimates. *Journal of Quantitative Analysis in Sports, 4*(2). https://doi.org/10.2202/1559-0410.1108

Albert, J. (2016). Improved component predictions of batting and pitching measures. *Journal of Quantitative Analysis in Sports, 12*(2), 73–85. https://doi.org/10.1515/jqas-2015-0063

Albert, J., & Bennett, J. (2003). *Curve ball: Baseball, statistics, and the role of chance in the game*. Copernicus Books.

Baumer, B. S. (2009). Using simulation to estimate the impact of baserunning ability in baseball. *Journal of Quantitative Analysis in Sports, 5*(2). https://doi.org/10.2202/1559-0410.1174

Bishop, C. M. (2006). *Pattern recognition and machine learning*. Springer.

Brill, R. S., Deshpande, S. K., & Wyner, A. J. (2023). A Bayesian analysis of the time through the order penalty in baseball. *Journal of Quantitative Analysis in Sports, 19*(4), 245–262. https://doi.org/10.1515/jqas-2022-0116

Bukiet, B., Harold, E. R., & Palacios, J. L. (1997). A Markov chain approach to baseball. *Operations Research, 45*(1), 14–23. https://doi.org/10.1287/opre.45.1.14

Deshpande, S. K., & Wyner, A. (2017). A hierarchical Bayesian model of pitch framing. *Journal of Quantitative Analysis in Sports, 13*(3), 95–112. https://doi.org/10.1515/jqas-2017-0027

Ding, L., Cribben, I., Ingolfsson, A., & Tran, M. (2025). Do NHL goalies get hot in the playoffs? A multilevel logistic regression analysis. *Journal of Sports Analytics, 11*. https://doi.org/10.1177/22150218251314737

Friedman, J. H. (2001). Greedy function approximation: A gradient boosting machine. *Annals of Statistics, 29*(5), 1189–1232. https://doi.org/10.1214/aos/1013203451

Gelman, A., Carlin, J. B., Stern, H. S., Dunson, D. B., Vehtari, A., & Rubin, D. B. (2013). *Bayesian data analysis* (3rd ed.). Chapman and Hall/CRC.

Gray, R. (2017, June 24). Pitch tunneling & perceptually equivalent pitches. *The Perception & Action Podcast*. https://perceptionaction.com/pitchtunnels/

James, G., Witten, D., Hastie, T., & Tibshirani, R. (2021). *An introduction to statistical learning* (2nd ed.). Springer.

Kaplan, D. (2006). A variance decomposition of individual offensive baseball performance. *Journal of Quantitative Analysis in Sports, 2*(3). https://doi.org/10.2202/1559-0410.1035

Lambert, J. (2024, May 31). Revisiting Stuff+: An update on Driveline's methodology to quantifying pitch design. *Driveline Baseball*. https://drivelinebaseball.com/blogs/blog/revisiting-stuff-plus

Lundberg, S. M., & Lee, S.-I. (2017). A unified approach to interpreting model predictions. *Advances in Neural Information Processing Systems, 30*, 4765–4774.

Piette, J., & Jensen, S. T. (2012). Estimating fielding ability in baseball players over time. *Journal of Quantitative Analysis in Sports, 8*(3). https://doi.org/10.1515/1559-0410.1463

Roegele, J. (2014, November 24). The effects of pitch sequencing. *The Hardball Times*. https://tht.fangraphs.com/the-effects-of-pitch-sequencing/

Romer, D. (2006). Do firms maximize? Evidence from professional football. *Journal of Political Economy, 114*(2), 340–365. https://doi.org/10.1086/501171

Sayre, B., Pavlidis, H., Judge, J., & Long, J. (2017, January 24). Prospectus feature: Introducing pitch tunnels. *Baseball Prospectus*. https://www.baseballprospectus.com/news/article/31030/prospectus-feature-introducing-pitch-tunnels/

Spilker, F., & Ötting, M. (2024). No cheering in the background? Individual performance in professional darts during COVID-19. *Journal of Quantitative Analysis in Sports, 20*(3), 219–234. https://doi.org/10.1515/jqas-2022-0036

Steeger, G. M., Dulin, J. L., & Gonzalez, G. O. (2021). Winning and losing streaks in the National Hockey League: Are teams experiencing momentum or are games a sequence of random events? *Journal of Quantitative Analysis in Sports, 17*(3), 155–170. https://doi.org/10.1515/jqas-2020-0077

Stern, H. (1991). On the probability of winning a football game. *The American Statistician, 45*(3), 179–183. https://doi.org/10.1080/00031305.1991.10475798

Sutton-Brown, S. (2025, January 17). Introducing BP's new arsenal metrics. *Baseball Prospectus*. https://www.baseballprospectus.com/news/article/96026/introducing-new-arsenal-metrics/

Weinstein-Gould, J. (2009). Keeping the hitter off balance: Mixed strategies in baseball. *Journal of Quantitative Analysis in Sports, 5*(2). https://doi.org/10.2202/1559-0410.1173